\documentstyle[12pt,aasms]{article}

\begin{document}


\title{A Dust Lane in the Radio galaxy 3C270}

\author{Ashish Mahabal, Ajit Kembhavi \\
        Inter-University Centre for Astronomy and Astrophysics \\
        Post Bag 4, Ganeshkhind, Pune 411 007, India,\\
\vskip0.1in
        K. P. Singh, P. N. Bhat \\
        Tata Institute of Fundamental Research \\
        Homi Bhabha Road
        Bombay 400 005, India
\and
        T. P. Prabhu \\
        Indian Institute of Astrophysics \\
        Bangalore 560 034, India
}

\begin{abstract}

We present broad band surface photometry of the radio galaxy 3C270
(NGC~4261).  We find a distinct dust lane in the $V-R$ image of the
galaxy, and determine its orientation and size.  We use the major
axis profile of the galaxy to estimate the optical depth of the dust
lane, and discuss the significance of the lane to the shape of the galaxy.
\vskip0.4in

\noindent {\it Subject headings:}
Galaxies: active; Galaxies: radio; Galaxies: individual; Galaxies:
interstellar medium; Galaxies: photometry; Galaxies: dust lane

\end{abstract}

\eject


\section{Introduction}

3C270 (NGC 4261) is a radio galaxy with a bright compact radio nucleus and
prominent jets and lobes.  While at first sight it appears to be  a
morphologically simple
elliptical galaxy of type E2, closer examination reveals many
interesting features.  A dust lane roughly along the apparent major axis
 of the galaxy was reported by M\H ollenhoff \& Bender (1987), while
others have failed to confirm its presence (e.g. Kormendy \& Stauffer 1987;
 Peletier 1990). On a much smaller angular scale, recent observations with
 the Planetary Camera on the Hubble Space Telescope (Jaffe et al. 1993)
 have led to the detection of an absorbing disk of radius
 $\sim2\times10^{20}\rm\,cm$ which is made up of cool dust and gas.  The
plane of the disk is
nearly normal to the radio jets, which are oriented
close to the apparent minor axis of the galaxy.  Long slit spectra
(Davies \& Birkinshaw 1986) have shown that the galaxy rotates round an
apparent axis which lies only $6^\circ \pm 4^\circ$ from the
apparent major axis, which suggests that the galaxy is prolate or close
to prolate.

In this paper we use $V$ and $R$ images of NGC 4261 to confirm the
existence of the dust lane
first noted by M\H ollenhoff \& Bender (1987).  We demonstrate the
existence of the lane in the combined $V-R$ colour image of the galaxy,
determine its size and orientation, and estimate its optical
depth from the colour excess relative to the region immediately outside
the dust lane.  We then independently estimate the optical depth of the
dust lane by examining the deviation of the surface brightness profile
of the galaxy from de Vaucouleurs' law. We end with comments on the
contribution of
the dust lane to the observed boxiness of galaxy isophotes and the
significance of the orientation of the lane to the radio
structure and intrinsic shape of the galaxy.


\section{Observation and Data Reduction}

NGC 4261 was observed at the prime focus $(f/3.5)$ of the $2.3\rm\,m$
{\it Vainu Bappu Telescope} at Kavalur on the night of March 3, 1992
as part of a programme of broad band imaging of radio galaxies.  The
 detector used was a GEC CCD with $576\times385$ pixels, each of
$22\times22\,\mu\rm m^2$ in size,
corresponding to $0.56\times0.56\rm\,arcsec^2$ at the prime focus.  We
obtained two  $600\rm\,s$ exposure frames in  $V$ and two   $200\rm\,s$
frames in $R$,  a number of twilight sky flats and bias frames spread
 throughout the night.  We observed standard stars in the {\it dipper
 asterism} region of the open cluster M 67 for calibration; the procedure
 adopted has been described in detail by Anupama et al. (1994).

We followed the standard procedure for bias subtraction and flat
fielding, and performed sky subtraction for each frame using a mean sky
 value estimated from pixels in the four corners of the frame, all of
which were unaffected by the galaxy.  We
combined the two frames in each filter  into single $V$ and $R$ frames
after performing translations and rotations so that the images were
aligned to better than $0.1$ pixel.  Since the night was not photometric,
 we did not apply any extinction corrections.  Tasks from IRAF\footnote{IRAF
is distributed by the National Optical Astronomy Observatories, which
is operated by the Association of Universities, Inc. (AURA) under cooperative
agreement with the National Science Foundation.} and STSDAS\footnote{The
 Space Telescope Science Data Analysis System STSDAS is distributed by
 the Space Telescope Science Institute.} were used
for all the data reduction and analysis.

Boxiness is clearly discernible in the isophotes of NGC 4261, which
nevertheless can be very well approximated by smooth ellipses.
We fitted ellipses to the isophotes
using routines in STSDAS, which are based on the method outlined by
Jedrzejewski
(1987).  The semi-major axes of successive fitted ellipses were larger by a
factor of $1.1$, and the fit extended from the centre to a semi-major axis
length
of $60''$, where the error in the surface brightness reaches
$\sim 0.1\rm\,magnitude$.  The semi-major axis brightness profile, ellipticity
and position angle of the fitted ellipses as functions of semi-major
axis length $a$ for the $V$ filter are shown in Figure 1.
These parameters show similar behaviour in $R$. The rather large error-bars
in case of ellipticity and position angle could be due to genuine departure of
isophots from true ellipses.
 The deviation of an
isophote from a perfect ellipse is estimated by expanding the
difference in intensity between the isophote and the corresponding
fitted ellipse as
a Fourier series in the eccentric anomaly along the isophote.  When boxiness is
present, the coefficient $b4$ of the $\cos(4\phi)$ term is negative.  We have
shown in Figure 1 the  ratio $b4/a$ as a function of $a$ for images in the $V$
filter. All the distributions in Figure 1 are in good agreement with the values
obtained by M\H ollenhoff \& Bender (1987).


\section{Colour Image}

The putative dust lane in NGC 4261 is too faint to be seen in a single direct
image, and the best place to look for it is a 2-colour image of the galaxy.
Since we do not have a $B$ frame with good signal-to-noise ratio, we
investigate
the $V-R$ image. The difference in the full width at half maximum (FWHM)
in the point spread function (PSF) for the $V$ and $R$ images is
 $\sim 0''.06\rm$,
compared to the FWHM of $\sim 2''.4\rm$. We therefore ignore the difference
in further analysis. A narrow elongated structure is clearly seen in the $V-R$
image (Plate~1), oriented close to the apparent major axis of the galaxy. We
show below that the feature is redder than its surroundings, and interpret it
as a dust lane. In order to improve the appearance of the lane, we have also
constructed a modified colour image adopting the procedure recommended by
Sparks et al. (1985), which uses a smooth version of the higher wavelength
image.  This enhances the higher spatial frequency structures without
affecting large scale colour features, and reduces noise in the 2-colour
image as one of the components is smoothed.  For this purpose we first fitted
ellipses
to the $R$ image following the procedure described in the previous section,
then obtained a smooth image $R_{smooth}$ by interpolating between the
isophotes
using the ellipse parameters, and finally the modified colour image,
$V-R_{smooth}$,
which is shown in Plate 2. The elongated structure in the $V-R$ image is
 now more obvious than in Plate 1. The feature is also seen if the smooth
image is obtained with the inclusion of higher harmonics which account for
the boxiness of the isophotes. We wish to emphasize that all calculations
presented
 in this paper are based on the original, unsmoothed $V$,~$R$ and
 $V-R$ images.

A cut across the dust lane in the $V-R$ image, averaged over ten
rows around the galaxy centre, is shown in Figure~2. Even though the
 instrumental
colour profile in the figure is noisy in the outer parts, it is clear that
there is an overall shift to the blue as one moves towards the centre
of the galaxy.  A region obviously  more red in
colour than its immediate surroundings is seen straddling
the centre. The red colour persists over the narrow feature,
and we interpret it to be a thin disk of dust seen nearly edge on and
appearing as a dust lane.

{}From inspection of Plate 1, the extent of the dust lane is found
to be $\sim21\times6\rm\,arcsec^2$.  We have taken several $5$-pixel wide
cuts across the dust lane, and assuming that the maximum in the $V-R$ profile
in such a strip corresponds to the centre of the lane for that strip,
obtained the coordinates of the centre along its length.  A linear
least square fit to these points then provides a straight
line representation of the dust lane.  This is shown in Figure~3, along
with the direction of the major axis obtained from the ellipse fits,
and the direction of the large scale radio jet as given by Birkinshaw
\& Davies (1985).  The dust lane makes an angle of $9^{\circ}\pm1^{\circ}$
with the major axis, and an angle of $97^{\circ}\pm1^{\circ}$ with the radio
jets.

We have estimated the colour excess in the dust lane by comparing the colour in
it with that of the surroundings.  For this purpose we used a circular
aperture with radius $1''.68$ ($3\rm\,pixels$)  on
the $V$ and $R$ frames separately to extract magnitudes along the
lane and a region along the minor axis $10''$ from the centre,
which is free of dust.  The mean of the colour excess obtained in this manner
is
	$$E(V-R) = (V-R)_{dust}-(V-R)_{galaxy} \simeq 0.05 \pm 0.01.$$

\noindent Assuming that the composition of the dust in the disk is similar to
 that of dust in our galaxy, and using standard relations from Savage \&
Mathius
(1979), it follows that $E(B-V)=0.06$, $\tau_{V} = 0.17$ and
$\tau_{R} = 0.08$, where $\tau$ is the optical depth.
Because these numbers are small the dust lane is not immediately apparent
 in the individual direct images. However, the dust does affect the
luminosity profile, and we explore that aspect in the next section.

The optical depth obtained requires a correction since the dust disk obscures
 only those stars which are behind it, and not the ones that are between the
disk and the observer. It can be shown (Brosch et al. 1990) that if the
fraction of stars obscured is $f$, the relation between the observed
 extinction $A_{\lambda}^{obs}$ and the optical depth $\tau_{\lambda}$ is

$$A_{\lambda}^{obs}=-2.5\log[1+f(e^{-\tau_{\lambda}}-1)].$$

\noindent
 For $\tau_{\lambda}\ll1$,
 the extinction is given by $A_{\lambda}^{obs}\simeq1.09f\tau_{\lambda}$, which
 may be compared to the usual relation $A_{\lambda}\simeq1.09\tau_{\lambda}$
 obtained by assuming that all the stars are obscured.
 The value of $f$ depends on the geometry and orientation of the disk.
 When the disk is inclined to the line of sight, as in the present case,
 it is expected that the near side of the disk is darker than the far side
\footnote{We wish to thank an anonymous referee for pointing this out to us.},
 and this could be used to fix the inclination when better data is available.

Following Burstein \& Heiles (1978), the column density of neutral
hydrogen and the total neutral hydrogen content in the dust disk can be
 obtained from

$$N(H) = 5.8\times10^{21} E(B-V){\rm atoms\ cm}^{-2}$$

\noindent and

$$N^{H}_{tot}= D^2 \int (dust) N(H) d\Omega.$$

\noindent Using $E(B-V)=0.06$ and $D=14.7 Mpc$ we get
$N^{H}_{tot} = 1.8 \times 10^{63}$,
i.e. $M^{H}_{tot} = 1.7\times10^{6} M_{\odot}.$ $H$ here refers to the
neutral hydrogen. Assuming that the dust
to gas mass ratio is $\sim10^{-2}$ as in our galaxy, we get the dust
mass $M_d = 1.5\times10^4 M_{ \odot }$, which is a lower limit since
a faint disk can extend beyond the confines presently detected. The
factor $f$ discussed above appears in the denominator of the integral
defining $N^{H}_{tot}$. Therefore, since $f<1$, the actual neutral
hydrogen content is larger than the estimate.


\section{Effect of Dust on the Luminosity Profile}

It is known that de Vaucouleurs' $r^{1/4}$ law describes very well the
observed brightness distribution for elliptical galaxies within
$0.1r_e \le a \le 1.5r_e$ (e.g. Burkert 1993), where $a$ and $r_e$
are the semi-major axis distance and effective radius respectively.
The lower limit is mainly
due to seeing effects. Following an iterative fitting procedure,
Burkert finds that for NGC~4261,
$r_e=34.5''$, so that de Vaucouleurs' law should hold
at least upto  $\sim3''.5$.
This procedure neglects the presence of the dust, which is
expected to lead to departures from the law because of the
absorption well above $0.1r_e$.  The distribution of
the surface brightness of NGC~4261 as a function of $r^{1/4}$, where $a$ is
in arcsecond is shown in Fig. 1.  It is clear
from the linear part of the curve for  $a > 11''$ that de Vaucouleurs'
law provides an excellent fit in this region.  The departure from a
straight line towards the centre is due to seeing, absorption and any real
deprtures from the law at small radii.

We have  fit de Vaucouleurs' law to the $V$ and $R$
profiles after excluding the inner $11''$.  For the fit,
a model galaxy was generated with assumed values of the effective radius
$r_e$ and central surface brightness  and using the observed distribution of
ellipticity.  The model was convolved with the point spread function (psf)
obtained from stars in
the frame, and the major axis profile was generated and compared
with the observed profile. Parameters of the model were determined using
the method of least squares.  The effective radius obtained in
this manner is $35''.8$.  The best fit profile together with
the observed surface brightness obtained from the ellipse fits described
in Section~2 is shown in Figure~4, and the agreement between the
two in the region used in the fit is seen to be excellent. If all points
of the observed profile are used in the fit, the effective radius $r_e$
obtained is $42''.6$, which is in agreement with the
value obtained by Peletier et al. (1990), but is an overestimate because
of the neglect of absorption.

The extrapolation of the best fit profile to the region of the dust disk lies
above the
observed points.  If this is attributed to the absorption due to dust,
which is certainly valid for $a \ge 3''.5$,
the difference $\Delta V$ in magnitude between the fit and the observed surface
brightness directly provides
a measure of the extinction, with optical depth $\tau_V=\Delta V/1.086$.
The optical depth obtained in this manner for the $V$ as well as
the  $R$ filter is shown in Figure~5.  There are two points to be
noted here : (1) The optical depth is independent of any
assumptions regarding the properties of the dust, and therefore can be
used to examine the validity of the assumptions made in Section~3 in
estimating optical depth from the colour excess . (2) The observed surface
brightness used in the determination of the extinction is obtained as a
result of averaging over the best fit ellipses, which
improves the signal-to-noise ratio and provides a smooth representation of the
dependence of the optical depth on the semi-major axis length.

For $8'' \le a \le 11''$ some points along  the fitted ellipses lie outside the
region
covered by dust, and extinction obtained  from the mean intensity as
described above could be an underestimate.  To check the magnitude of this
effect, we  have taken intensity cuts through
the centre which extended only through the  region covered by dust. These were
then  averaged and an intensity profile obtained. We found
that this profile differed at most by 0.01 magnitude from the profile
obtained using the fitted ellipses.  We therefore use the latter over the
entire
dust region.

The mean value of the optical depth over the dust region obtained using
the above method separately for the two filters is $\tau_{v}=0.19\pm0.01$ and
$\tau_{r}=0.10\pm0.01$, with the indicated error being primarily due
to the uncertainty in deciding the extent of the dust obscured region.
The optical depth here agrees within errors with the value in
Section~3, which justifies the assumption made there that the dust
in the disk is similar to the dust in our galaxy.

It is possible to look for departures from de Vaucouleurs' law
in the image rather than in the radial surface brightness profiles.
For this purpose, using the best fit de Vaucouleurs' profile outside
the putative dust region in the V filter, we have generated a 2-D model,
extrapolated it to the region covered by the dust and convolved the
whole with the point spread function. This provides a dust free
representation $V_{df}$ of the galaxy. The residual image $V-V_{df}$
shows features similar to $V-R_{smooth}$, with some variations which
arise because of the differences in model generation and smoothing
used in the two cases. The similarity in the residual and colour images
makes it reasonable to assume that de Vaucouleurs' law modified by dust
absorption provides a reasonable description of the inner region of
NGC~4261.

It is expected that the optical depth in $V$ is higher than that in $R$.
 However in Figure~5 it is seen that $\tau_{V}\simeq\tau_{R}$ for $a\ge4''$
and
for $a<4''$ there is  decrease in
optical depth $\tau_{R}$ while $\tau_{V}$ continues to rise.
Literally taken, this would mean increased red light towards the center.
Some ellipticals are known to have red nuclei (e.g. Sparks et al. 1985) and
red cores (e.g. Carter et al. 1983), but good spectroscopic data and
observations in the $B$ band under better seeing conditions will
 be necessary to confirm these trends here,
as well as the similarity of the dust
 in NGC 4261 and in our galaxy. Dust in the two galaxies could have
significantly different properties because of different origin and
environment. The active galactic nucleus in the radio galaxy could also
 have some effect on the dust (Begelman 1985; Shanbhag \& Kembhavi 1988).
 The results at small radii would of course be affected by any colour dependent
 departures from de Vaucouleurs' law.


\section{Dust and Isophote Boxiness}

The dust lane reduces the surface brightness at every point along
its extent, which results in the isophotes appearing to be pulled inwards.
 When the dust
lane is oriented along the major axis, this produces the
appearance of boxiness, with the coefficient $b4$ of the $\cos(4\phi)$ term
in the angular dependence of the differential intensity along the best fit
 ellipses becoming negative. To
examine the magnitude of this effect, we have generated an elliptical
galaxy with the central surface brightness and effective radius obtained
for the best fit model in Section~3, using tasks from IRAF that allow the
 introduction of Poisson shot noise and characteristics of the CCD.  After
convolving the
model with the observed psf, we have  placed in it a dust lane
which has the extent and optical depth observed in NGC~4261 in the $V$ band.
We have then fitted ellipses
to the isophotes, and obtained $b4/a$, which is shown in
Figure~6.  The magnitude of $b4/a$ is seen to be similar to the
observed value where dust is present, and the circularizing effect of the
seeing is not dominant.  However the boxiness is colour dependent, unlike
in the observed case, and reduces to insignificant values as soon as the
dust is left behind.  Nevertheless it is clear that  observed isophotal
boxiness could have a contribution from dust and could in fact be used as
a diagnostic of major axis dust lanes which are too faint to be seen directly.
 In
the same manner, dust lanes oriented in directions other than the apparent
major axis will produce changes in isophotal shape which can be traced in
the various Fourier coefficients.


\section{Discussion}

The shape and orientation of a dust disk in an elliptical galaxy
is determined by the allowed trajectories of the dust particles,
which depend on the shape and the overall rotation of the
galaxy.  NGC 4261 shows rotation around the apparent major axis (Davies \&
Birkinshaw 1986) and therefore it cannot be an oblate spheroid (Binney 1985).
Davies and Birkinshaw have argued from the kinematical data and
modelling that the galaxy is prolate or triaxial and nearly
prolate.  In such a configuration, orbits in the equatorial
plane (which is normal to the long axis) are stable, but these
are not appropriate to describe the  dust disk in NGC 4261, because the
dust lane is observed to be aligned close to the major axis. However there is
another class of orbits which is stable (see Kormendy \& Djorgovoski
(1989) and de Zeeuw \& Franx (1991) for a  review of the possibilities
and references) and has
different orientations at different radii which may be more appropriate
to the present case.  At small radii these orbits are polar, i.e. they
lie in a plane containing the long and short axes; at larger
radii the orbits are equatorial and skew at intermediate
radii.  The small scale absorption disk discovered using the Hubble Space
Telescope (Jaffe et al. 1993) is inclined at an angle of $\sim64^{\circ}$
to the plane of the sky, while the inclination of the larger disk is
$\sim75^{\circ}$.  The two disks are to be viewed as the warped parts of
a single disk, with the warping being due to the complex nature of
the orbits, and the possible change in the shape of the galaxy from prolate
to oblate towards the inner region, which will change the plane of stable
 orbits. The relative orientation of the different
parts of the disk are also dependent on the extent to which the parts have
settled in their final orbits, and projection effects which again depend
on the shape of the galaxy and the direction of the line of sight.  Higher
signal-to-noise observations under better seeing conditions as well as
spectroscopic data will be able to provide further information about the
extent and shape of the dust disk which can be used in modelling the galaxy.

The radio jet in NGC 4261 is oriented almost normal to the dust disk,
which hints at a possible connection between the two.  The direction of
the angular momentum of the disk changes somewhat from region to region as the
disk is warped. It would be reasonable for the direction of the
base of the jet to be determined by the innermost parts of the disk, and
precession of the disk would lead to a change in the direction of the jet.
However the large scale jet may overall appear to be normal to the large
scale disk. Based on the statistical analysis of radio jet directions
(e.g. Palimaka et al. 1979; Kapahi \& Saikia 1982),
it has been argued that the jets are oriented along the
apparent minor axis of ellipticals.  When the galaxy is an oblate spheroid,
the apparent minor axis always coincides with the projection of
the short axis. The observation then means that jets are preferentially
emitted close to the short axis. This coincidence could be related to the
fact that in non-rotating or slowly rotating oblate spheroids the equatorial
plane has stable orbits, so that the accretion disk is situated in this
plane or close to it.  When the galaxy is prolate, triaxial or
rotating, the situation is more complex, and as mentioned above a variety of
stable orbits is possible.  If the jet direction is again taken to be
determined
by a disk, the relation between the jet direction and the apparent minor axis
will depend on the the shape of the galaxy, the initial conditions of
dust formation and the viewing direction. One should also expect to find a disk
at least in those ellipticals which are powerful radio galaxies with
well collimated jets.  As in the case of NGC 4261, these disks may be
too faint to be observed directly, but it would not be difficult to spot
them using the techniques we have discussed above.  If ellipticals are
the results of mergers or cannibalism, one may well find dust disks in
all of them.


\section{Conclusions}

The main conclusions of the present paper are the following :

\begin{enumerate}

\item {The radio galaxy NGC 4261 contains a dust lane with dimensions
$\sim21\times6\rm\,arcsec^2$, oriented close to the apparent major axis of the
galaxy.  The dust lane can be interpreted as the projection of a dust disk
with inclination angle $\sim75^{\circ}$ to the plane of the sky.  The observed
colour excess is $E(V - R) = 0.05$.  Assuming dust properties similar to those
in our galaxy, the optical depth obtained from $A_{\lambda}\simeq
1.09\tau_{\lambda}$ is $\tau_{V} = 0.17$ and
$\tau_{R} = 0.08$.} The actual optical depth could be a factor $\sim$2 higher
 than this value.

\item {The optical depth can be estimated directly from departures of the
surface brightness profile from de Vaucouleurs' law, and has values close
to those obtained from the colour excess, confirming that the dust is similar
to that in our galaxy.}

\item {Absorption due to a major axis dust lane can produce boxiness
which is colour dependent.}

\end{enumerate}


\section{Acknowledgements}
The authors wish to thank the staff of the Vainu Bappu Observatory, Kavalur
for help with the observations, and Professor D. Lynden-Bell,
Dr. G. C. Anupama, Dr.~S.~K.~Pandey for many enlightening discussions.

\eject


\eject


\noindent {\bf Figure Captions}

\vskip0.1in
\noindent {\bf Figure 1 : } Semi-major axis profiles in the $V$ band of
surface brightness, ellipticity, position angle and the {\it boxiness
 coefficient} $b4/a$. The surface brightness is shown as a function of
 $a^{1/4}$, where $a$ is the semi-major axis length, of the isophotes.

\vskip0.15in
\noindent {\bf Figure 2 : } A cut across the dust lane, averaged over
 10 rows around the galaxy centre. The instrumental colour $v-r$ is shown.
\vskip0.15in

\noindent{\bf Figure 3 : } A schematic representation of NGC 4261.
 The major axis lies within the narrow confines of the dust lane and
 the radio jet is almost perpendicular to it.
\vskip0.15in

\noindent {\bf Figure 4 : } Observed brightness profile along the
semi-major axis, and a de Vaucouleurs' law fit made using points beyond
the small vertical bar shown at $a=11''$. The fit is extrapolated to the
 region inward of the bar.
 Deviation of the fit from the observed
 points in the inner region is indicative of the extinction due to dust.
\vskip0.15in

\noindent {\bf Figure 5 : } Optical depth $\tau_{V}$ and $\tau_{R}$ as
 function of $a$. The value at a given $a$ is the mean over the best fit
 ellipse with that semi-major axis.
\vskip0.15in

\noindent {\bf Figure 6 : } The {\it boxiness coefficient} for a simulated
 elliptical galaxy generated using the parameters obtained for NGC 4261.
\vskip0.15in

\noindent {\bf Plate Captions}

\vskip0.1in
\noindent {\bf Plate 1 : } $V-R$ image of NGC 4261 in false colour.
 Each pixel is $0''.56.$

\vskip0.15in
\noindent {\bf Plate 2 : } $V-R_{smooth}$ image of NGC 4261 in false colour.
Each pixel is $0''.56.$



\begin{thebibliography}{}

\bibitem{} Anupama, G. C., Kembhavi, A. K., Elvis, M., \& Edelson, R.
1994, A\&AS, 103, 315

\bibitem{} Begelman, M. C. 1985, ApJ, 279, 492

\bibitem{} Binney, J. 1985, MNRAS, 212, 767

\bibitem{} Birkinshaw, M. \& Davies, R. L. 1985, ApJ, 291, 32

\bibitem{} Brosch, K., Almoznino, E., Grosbol, P., \& Greenberg, J. M.
1990, A\&A, 233, 341

\bibitem{} Burkert, A. 1993, A\&A, 278, 23

\bibitem{} Burstein, D. \& Heiles, C. 1978, ApJ, 225, 40

\bibitem{} Capetti, A., Macchetto, F., Sparks, W. B., \& Miley, G. K.
1994, A\&A, 289, 61

\bibitem{} Carter, D., Jorden, P. R., Thorne, D. J., Wall, J. V., \& Straede,
J. C.
1983, MNRAS, 205, 377

\bibitem{} Condon, J. J. \& Broderick, J. J. 1988, AJ, 96(1), 30

\bibitem{} Davies, R. L. \& Birkinshaw, M. 1986, ApJ, 303, L45

\bibitem{} de Zeeuw \& Franx, M. 1991, ARA\&A, 29, 239

\bibitem{} Fraix-Brunet, D., Golembek, D., Macchetto, F., Nieto, J. L.,
Lelievre, G., Perryman, M.~A.~C., \& Di  Serego Alighieri, S. 1991, AJ,
 101(1), 88

\bibitem{} Jaffe, W., Ford, H. C., Ferrarese, L., van den Bosch, F., \&
O'Connell, R. W. 1993, Nature, 364, 213

\bibitem{} Jedrzejewski, R. 1987, MNRAS, 226, 747

\bibitem{} Jenkins, C. R. 1981, MNRAS, 196, 987

\bibitem{} Jones, D. L., Sramek, R. A., \& Terzian, Y. 1981, ApJ, 246, 28

\bibitem{} Kapahi, V. K. \& Saikia, D. J. 1982, JA\&A, 3, 161

\bibitem{} Knapp, G. R., Bies, W. E., \& van Gorkom, J. H. 1990, AJ, 99(2), 476

\bibitem{} Kormendy, J. \& Djorgovski, S. 1989, ARA\&A, 27, 235

\bibitem{} Kormendy, J. \& Stauffer, J. 1987, IAU symp., 127, 405

\bibitem{} Lauer, T. 1985, MNRAS, 216, 429

\bibitem{} Maltby, P., Matthews, T. A., \& Moffet, A. T. 1963, ApJ, 137, 153

\bibitem{} M\H ollenhoff, C. \& Bender, R. 1987, A\&A, 174, 63

\bibitem{} M\H ollenhoff, C., Hummel, E., \& Bender, R. 1992, A\&A, 255, 35

\bibitem{} Palimaka, J. J., Bridle, A. H., Fomalont, E. B., \& Brandie, G. W.
1979, ApJ, 231, L7

\bibitem{} Peletier, R., Davies, R. L., Illingworth, G. D., Davis, L. E., \&
Cawson, M. 1990, AJ, \hspace{1cm} 100(4), 1091

\bibitem{} Savage, B. D. \& Mathis, J. S. 1979, ARA\&A, 17, 73

\bibitem{} Schweizer, F. 1979, ApJ, 233, 23

\bibitem{} Shanbhag, S., \& Kembhavi, A. K. 1988, ApJ, 334, 34

\bibitem{} Sparks, W. B., Wall, J. V., Thorne, D. J., Jorden, P. R., van Breda,
I. G., Rudd, P. J., \&
 Jorgensen, H. E. 1985, MNRAS, 217, 87

\end{thebibliography}
\end{document}